\documentclass[reprint,amsmath,amssymb,
aps,prl,lengthcheck,nobibnotes]{
revtex4}
\usepackage{graphicx}
\usepackage{dcolumn}
\usepackage{bm}
\usepackage{hyperref}
\usepackage[geometry]{ifsym}
\usepackage{color}
\usepackage{natbib}
\begin{document}


\title{New isomeric transition in $^{36}$Mg: Bridging the {\it N}=20 and {\it N}=28 islands of inversion}

%
%
\author{M. Madurga$^1$}
\author{J.M. Christie$^1$}\author{ Z. Xu$^1$} \author{R. Grzywacz$^{1,2}$} \author{A. Poves$^3$} \author{T. King$^1$}
\author{J.M. Allmond$^2$}
\author{ A. Chester$^4$} \author{I. Cox$^1$} \author{ J. Farr$^1$}
\author{I. Fletcher$^1$} \author{J. Heideman$^1$}\author{ D. Hoskins$^1$}\author{ A. Laminack$^2$}\author{ S. Liddick$^{4,5}$} \author{ S. Neupane$^1$} 
\author{ A.L. Richard$^{4}$}\altaffiliation[Current address: ] {Institute of Nuclear and Particle Physics and Department of Physics and Astronomy, Ohio University, Athens, Ohio 45701, USA.\\} \author{N. Shimizu$^6$} \author{P. Shuai$^{1,2}$}\altaffiliation[Current address: ]{Institute of Modern Physics, Chinese Academy of Sciences, Lanzhou, Gansu 730000, China}
\author{K. Siegl$^1$} \author{Y. Utsuno$^{7,8}$}
\author{P. Wagenknecht$^1$}\author{ R. Yokoyama$^1$} 
\affiliation{
$^{1}$Dept. of Physics and Astronomy, University of Tennessee, Knoxville, Tennessee 37996, USA\\
$^{2}$Physics Division, Oak Ridge National Laboratory, Oak Ridge, Tennessee 37830, USA\\
$^{3}$Departamento de F\'{\i}sica Te\'orica, and IFT UAM-CSIC, Universidad Aut\'onoma de Madrid, 28049, Madrid, Spain\\
$^{4}$National Superconducting Cyclotron Laboratory, Michigan State University, East Lansing, Michigan 48824, USA\\
$^{5}$Department of Chemistry, Michigan State University, East Lansing, Michigan 48824, USA\\
$^{6}$Center for Computational Sciences, University of Tsukuba, 1-1-1, Tennodai Tsukuba, Ibaraki 305-8577, Japan\\
$^{7}$Advanced Science Research Center, Japan Atomic Energy Agency, Tokai, Ibaraki 319-1195, Japan\\
$^{8}$Center for Nuclear Study, University of Tokyo, Hongo, Bunkyo-ku, Tokyo 113-0033, Japan}

\date{\today}

\begin{abstract}
We observed a new isomeric gamma transition at 168 keV in $^{36}$Mg, with a half-life of $T_{1/2}=90(^{+410}_{-50})$ ns. We propose that the observed transition de-excites a new 0$^+$ isomeric state at 833 keV and populates 
the previously known first 2$^+$ state. The existence of this isomer is consistent with the predictions of the large-scale shell model calculations of $^{36}$Mg using the { \sl sdpf-u-mix} interaction.
The observed excitation energy of the second 0$^+$ state is caused by the small energy separation between two prolate-deformed configurations where the intruder configuration corresponds to two neutron excitations from the {\it sd} to the {\it pf} shell.  Within this interpretation, $^{36}$Mg becomes the crossing point between nuclei in which ground state deformed/superdeformed configurations are caused by the dominance of {\it N}=20 intruders ($^{32,34}$Mg)  and nuclei where deformed configurations are associated with the breaking of the {\it N}=28 closure and a large occupancy of the $1p_{3/ 2}$ neutron orbit   ($^{38}$Mg and beyond). We found the lack of three-body monopole corrections in other effective interactions results in a predominance of {\it N}=20 intruder configurations past $^{38}$Mg incompatible with our observation. We conclude that $^{36}$Mg bridges the {\it N}=20 and {\it N}=28 islands of inversion, forming the so-called Big Island of Deformation.

\end{abstract}

\maketitle


The large shell gaps in nuclei with "magic" numbers for protons and neutrons emerge from the collective action of the strong forces mediated through pion exchange. However, most nuclei are non-magic, and many are deformed due to the effects of nucleon-nucleon correlations.
The surprising emergence of the so-called islands of inversion, where the nuclei with magic numbers are known to be deformed, was attributed to the dominating character of correlations due to the quenching of the shell gaps.\\ 
The Island of  Inversion centered around magnesium isotopes with neutron “magic” number {\it N}=20 has attracted considerable interest \cite{Poves1987,Warburton1990,Heyde1991,Fukunishi1992,Caurier2014} since its discovery \cite{Thibault1975}. Negative-parity intruder states ascribed to excitations involving multiple particle-hole configurations between {\it sd} to the {\it pf} orbitals indicate a sudden quenching of the {\it N}=20 shell closure. Nuclei inside the Island of Inversion are defined by having ground states dominated by such configurations \cite{Neyens2005}. Further, recent experimental \cite{Crawford2019} and theoretical \cite{Caurier2014} studies suggest that particle-hole configurations dominate ground states in this region of the chart of nuclei between the {\it N}=20 and the {\it N}=28 "magic" numbers.  This forms a so-called Big Island of Deformation, where both neutron "magic" numbers {\it N}=20 and {\it N}=28 disappear in the magnesium isotopic chain.The quenching of the {\it N}=20 and {\it N}=28 neutron gaps is driven by the diminishing effect of the isospin {\it T}=0 component of the tensor force as the proton-neutron ratio becomes more asymmetric \cite{Otsuka2005}. Recently developed interactions in the proton/neutron {\it sd-pf} valence space have had considerable success in reproducing the observed intruder and ground-state configurations of known Island of Inversion nuclei. Some examples are effective interactions such as {\sl sdpf-m} \cite{Utsuno1999}, which only includes the $0f_{7/ 2}$ and $1p_{3/ 2}$ neutron orbits and we shall disregard,
{\sl sdpf-u-mix} \cite{Caurier2014}, or the new interaction {\sl EEdf1}, developed from the chiral expansion  at N3LO.
\cite{Tsunoda2017}. 
As we shall see later the explicit three body global monopole term proposed with {\sl sdpf-u-mix} is crucial to produce the evolution of the {\it N}=20 neutron closure towards {\it N}=28 consistent with our observation.
Interestingly, both interactions predict differing microscopic interpretations of the merging of the {\it N}=20 and {\it N}=28 islands of inversion. In  {\sl EEdf1}  excited states crossing the {\it N}=20 shell closure are substantial in both islands of inversion. On the other hand, {\sl sdpf-u-mix} predicts the restoration of the {\it N}=20 shell closure at $^{40}$Mg.
postulating instead that deformation is driven exclusively by the breakdown of the {\it N}=28 subshell closure. There is currently no experimental data that can resolve these differing interpretations. Delineation of the boundaries of the islands of inversion towards the neutron drip-line is therefore essential to determine the disappearance and appearance of the {\it N}=20 and {\it N}=28 shell closures respectively \cite{Fossez2017}. Isomers, long lived excited states, offer an observable with which to track evolving nuclear properties as we study nuclei between shell closures. The half-life of an isomeric state is fully determined by the transition's energy and its electromagnetic transition probability, in turn defined by the wave-functions of the involved states. One such example are the so-called shape isomers, excited states arising from nuclear configurations of different shapes.  Low energy excited $0^+$ states corresponding to prolate(oblate) deformed configurations \cite{Hamilton1974} may become isomeric when decaying to the first excited $2^+$ state corresponding to the ground state band of different deformation.

As of the beginning of 2023, there is only one isomer confirmed and published in either neon or magnesium isotopes, the 0$^+_2$ state in $^{32}$Mg that decays to the 2$^+_1$ via a 172 keV transition with $T_{1/2}>$10 ns \cite{Wimmer2010,Elder2019}. Shell model calculations using the {\sl sdpf-u-mix} interaction \cite{Poves2016} 
produce a ground state that is a mixture of deformed ($2\hbar\omega$) and superdeformed ($4\hbar\omega$) configurations and an isomeric 0$^+$ state which is dominated by
superdeformed and spherical ($0\hbar\omega$) components \cite{Caurier2014}. Notice that {\sl sdpf-u-mix} is the only interaction that locates the isomer close to its experimental excitation energy.
In the same calculation, heavier magnesium isotopes were expected to strongly favor quadrupole components before transitioning to the {\it N}=28 Island of Inversion at $^{40}$Mg. This hypothesis is supported by the systematics of the first 2$^+$ states in $^{34,36,38}$Mg \cite{Doornenbal2013,Doornenbal2016,Crawford2014,Crawford2019} comparing well with calculations \cite{Nowacki2021}. 

In this work we present the observation of a new isomeric gamma transition at 168 keV in $^{36}$Mg. Based on the observation of a 665 keV gamma line, likely corresponding to the $2^+$ state in $^{36}$Mg, we propose it corresponds to a second 0$^+$ state feeding said first 2$^+$ state. The analysis of the time structure of 168 keV gamma-ray events following the ion implantation results in a half-life of  $T_{1/2}=90(^{+410}_{-50})_{stat}(\pm40)_{tran}(^{+800}_{-70})_{sys}$ ns ("tran" corresponds to the uncertainty due to the transit time A1900, see below). We present an interpretation of the nature of the new second 0$^{+}$ state and the evolution of intruder configurations in the magnesium isotopic chain from {\it N}=20 to {\it N}=28 using shell model with configuration interaction (SM-CI) calculations with the {\sl sdpf-u-mix} interaction \cite{Caurier2014}. Our calculations indicate the isomer naturally arises from gradually restoring the {\it N}=20 shell closure as the neutron $0f_{7/2}$ orbital is occupied towards the {\it N}=28 subshell closure. 

Our calculations of $^{36}$Mg the observed low energy of the second 0$^+$ state by the small energy separation between 0p-0h and 2p-2h configurations. This makes $^{36}$Mg the crossing point between dominant ground state 2p-2h/4p-4h configuration in $^{32,34}$Mg and dominant 0p-0h configuration in $^{38}$Mg. We conclude $^{36}$Mg is the bridge between the {\it N}=20 and {\it N}=28 islands of inversion.

\begin{figure}
\begin{center}
\includegraphics[width=3.4in]{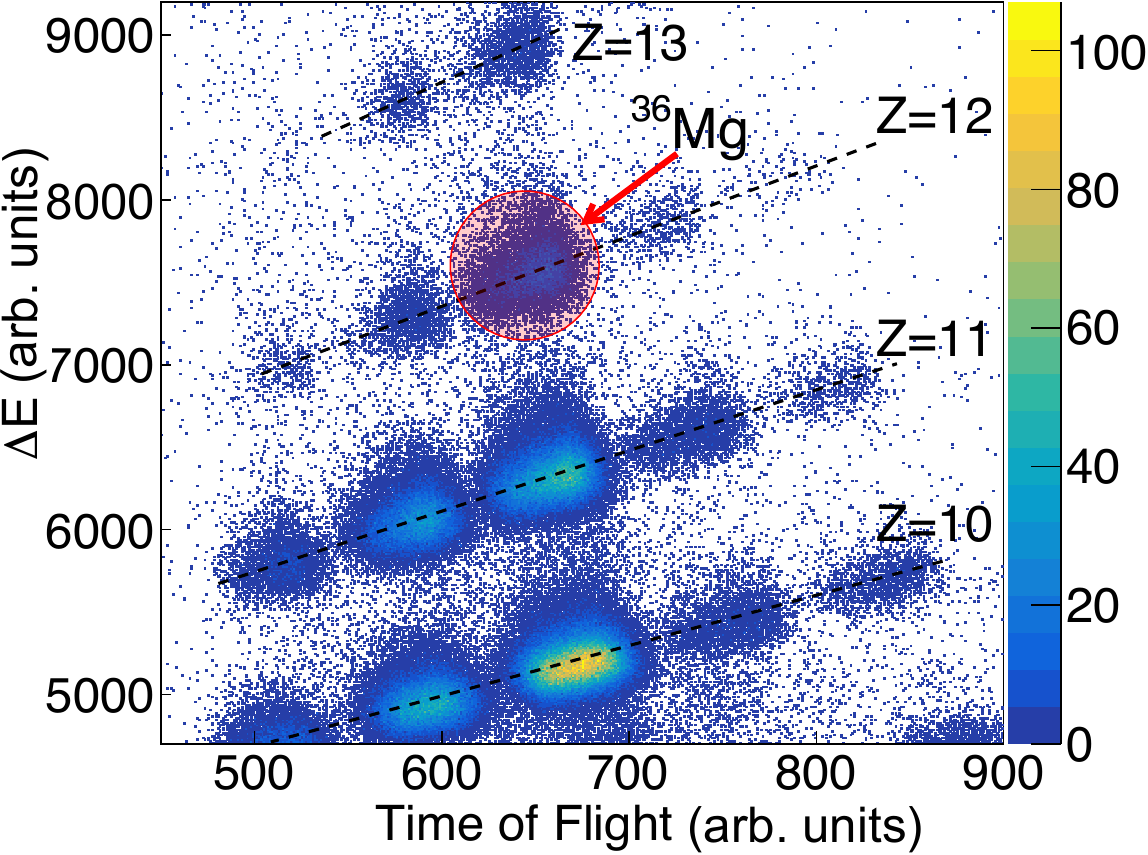}
\caption{Two dimensional energy loss ($\Delta$E) v. time of flight particle identification plot for all ion implants between {\it Z}=10 (bottom row) and {\it Z}=13 (top row). Magnesium-36 is highlighted by the red circle. We also searched for isomers in $^{25-29}$F isotopes (not shown). }
\label{fig1}
\end{center}
\end{figure}

{\it Experiment.} The experiment was performed at the National Superconducting Cyclotron Laboratory (NSCL) at Michigan State University. A $^{48}$Ca beam, 80 pnA average intensity at 140 MeV/u, was fragmented in a 846 mg/cm$^2$ thick Be target at the entrance of the fragment separator, A1900 \cite{Morrissey2003}, to produce the nuclei of interest,  
a "cocktail" beam consisting of isotopes from boron ({\it Z}=5) to aluminum ({\it Z}=13). 
In order to identify the different species, we measured the ion's time-of-flight between a scintillator located in the focal plane of A1900 and a Silicon detector (Si PIN) placed in front of our experimental setup. Combining with the energy loss in the Si PIN allowed us to perform particle identification (PID) in the beam, as shown in Fig. \ref{fig1}. We implanted the "cocktail" beam in a 12-mm thick YSO detector (Yttrium Orthosilicate, Y$_2$SiO$_5$) \cite{Yokoyama2019} allowing for recording energies and timestamps of ion implantation and beta-decay events.
The YSO detector was surrounded on one side by 48 VANDLE modules \cite{Peters2016} providing a total neutron detection efficiency of 11\% at 1 MeV. On the other side of the setup, three HPGe clovers from the CLARION array \cite{Gross2000} resulting in gamma detection of 1.3\% efficiency at 1 MeV.

\begin{figure}
\begin{center}
\includegraphics[width=3.4in]{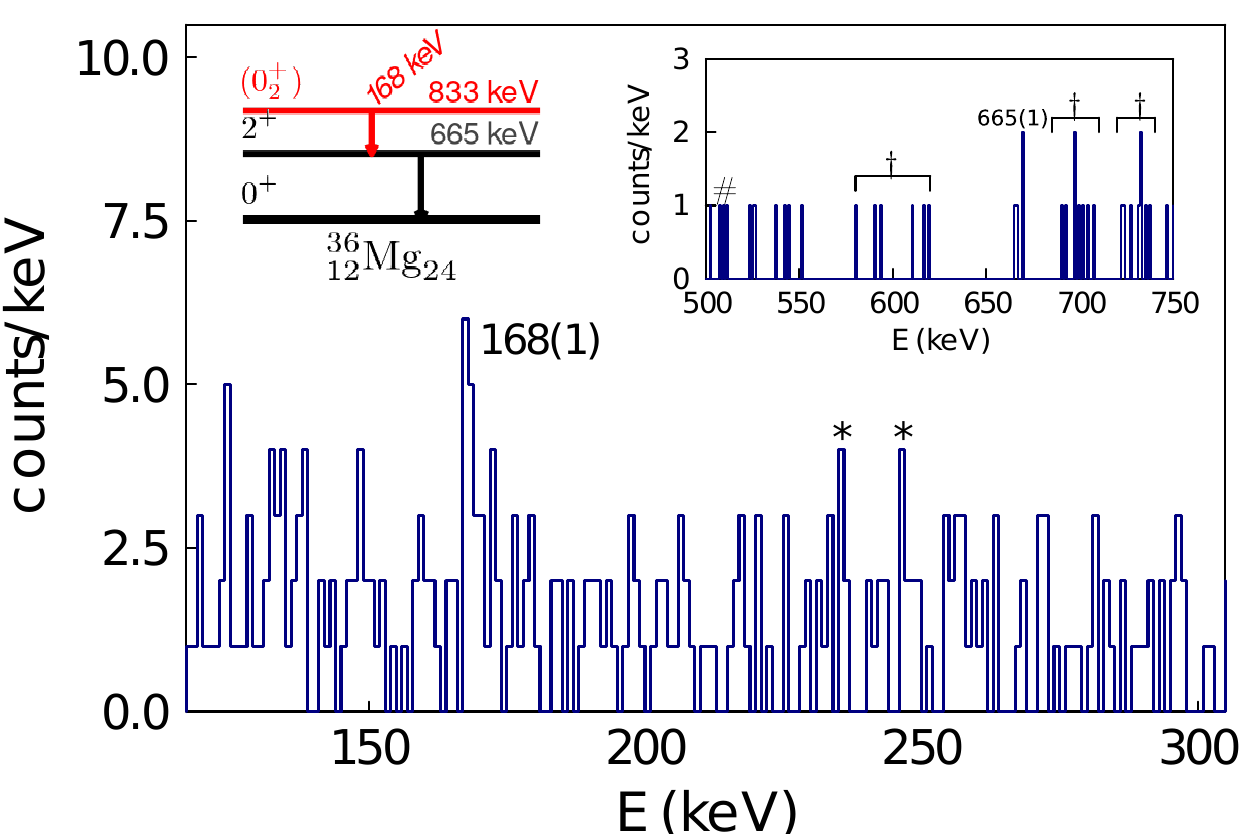}
\caption{Delayed gamma energy spectrum in coincidence with $^{36}$Mg implantation events. The most prominent line corresponds to the new isomeric transition at 168 keV ( [*] marks background lines).  The top left panel shows the $^{36}$Mg partial level scheme. The top right panel shows the gamma spectrum between 500 and 750 keV, including the $^{36}$Mg 665 keV transition and other background lines  (see text for details).}
\label{fig2}
\end{center}
\end{figure}

We searched for isomers in all Fluorine, Neon, Sodium, Magnesium, and Aluminum isotopes shown in Fig. \ref{fig1} by analyzing the gamma rays emitted between 40 ns and 500 ns after ion implantation, correlated to each individual isotope using the PID plot (Fig. \ref{fig1}). We excluded the first 40 ns in order to remove the Gaussian tail of the prompt implantation "flash". 
We did not identify isomeric transitions in any F, Ne, Na, Mg, or Al isotope except for $^{36}$Mg ($^{32}$Mg was outside of the separator acceptance in our experiment). In $^{36}$Mg, we observe a prominent gamma transition at 168 keV, see Fig. \ref{fig2}. The top right panel of Fig. \ref{fig2} shows the gamma spectrum between 500 and 750 keV. We marked several gamma lines ($\dagger$) corresponding to neutron inelastic scattering in common HPGe materials \cite{Baginova2018}, as well as the 511 keV line corresponding to positron annihilation ($\#$). We observe 4(2) counts (all errors in this section  are statistical at $1\sigma$ confidence level) in the energy region where the 665 keV  the de-excitation of the first excited state in $^{36}$Mg \cite{Gade2007,Doornenbal2013,Kobayashi2014,Michimasa2014,Doornenbal2016,Crawford2019} would be located.  The spin and parity of the first excited state of $^{36}$Mg  was confirmed to be 2$^+$ by the recent measurement of the quadrupole electromagnetic transition strength \cite{Doornenbal2016}. Since the state was not observed to be isomeric, we propose the isomeric state in $^{36}$Mg decays to the 2$^+$ state via emitting the 168 keV gamma ray. We calculated the number of counts we would observe if the new 168 keV line and the 665 keV line form a gamma cascade. We observe 8(4) counts in the 168 keV peak above background. Using efficiencies of 1.8\% at 168 keV and 1.5\% at 665 keV
we expect 6(3) counts at 665 keV, compatible with the observed 4(2) counts. Imposing total event multiplicity one and using the complete dataset with no isotope selection, we observe nothing but Compton background in the 640 to 680 keV region. To further validate our hypothesis, the presence of a 665 keV line, we performed a statistical study using the Monte-Carlo method, obtaining a confidence level of 2.8$\sigma$.
Assuming the two lines are in coincidence, as presented by the evidence above, we propose that the 168 keV isomer de-excites a new 833 keV state directly to the known 665 2$^+_1$ state in $^{36}$Mg (top left inset in Fig. \ref{fig2}). The spin parity of this new state cannot be directly measured in this experiment. However, we can identify possible candidates and rule out impossible combinations.  Given the strong evidence for the first 2$^{+}$  to correspond to the first excited state of a prolate rotational band \cite{Doornenbal2016}, any positive parity member of the band would not be isomeric. We can also rule out negative parity states of spin higher than 0 and lower than 4, as they would decay to the 2$^+$ via an E1 transition, typically too fast to be isomeric. This leaves $0^{+,-}$, and 4$^-$ and higher as the best candidate spin parities. Provided the state is at $833$ keV, it would be below the pairing gap, therefore making it very unlikely to be a negative parity state. We conclude $0^+$ is the most likely spin parity, corresponding well to the the other known isomer in neutron rich magnesium isotopes (the second 0$^+$ state in $^{32}$Mg \cite{Wimmer2010}). 

\begin{figure}
\begin{center}
\includegraphics[width=3.3in]{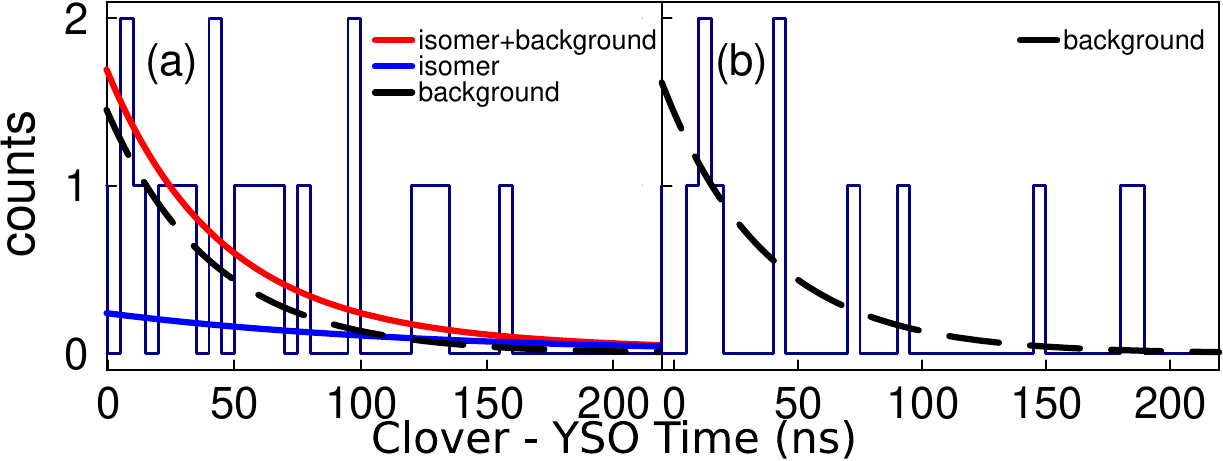}
\caption{(Left) Time distribution of gamma activity after ion implantation gated on the 168 keV photopeak with background (black-dashed, see right panel), isomer (blue) and combined (red) fits overlayed. (Right) Distribution of gamma events gated on the Compton background surrounding the photopeak.}
\label{fig3}
\end{center}
\end{figure}

We performed a log likelihood analysis of the gamma activity after ion implantation to determine the isomer's half-life, see Fig. \ref{fig3}. The left panel shows the time distribution of gamma events after ion implantation for the photopeak gate (167 to 169 keV). The right panel shows the time distribution of background events (166$<$E$_{\gamma}$$<$167 keV and 169$<$E$_{\gamma}$$<$170 keV). 
First, in order to estimate the component arising from the tail of the Gaussian distribution of the ion implantation Bremsstrahlung flash, we fitted the time distribution of the background gate to an exponential function (Fig. \ref{fig3}b). Then, we constructed a double exponential distribution, corresponding to photopeak and background combined, 50\% each as per the gamma energy spectrum estimate. We fitted the photopeak gate, obtaining $T_{1/2}$=$90(^{+410}_{-50})$ ns. We also studied the systematic uncertainty due to the shape of the tail of the implantation flash. In order to progressively remove the background tail, we performed fits to samples starting at increasingly later times, between 50 ns to 100 ns. Finally, we calculated the shortest observable half-life considering the transit time (500 ns) in A1900, and assuming an isomer population of 10\% (or a larger population of 40\%) in the fragmentation reaction producing $^{36}$Mg, obtaining 130$\pm40$ (90$\pm30$ for 40\% isomer population) ns. Given the statistical constraints due to the size of our sample (3$\sigma$ statistical uncertainty marked as "stat" in parenthesis), the limits imposed by the transit time in A1900 (uncertainty marked in parenthesis with "tran"), and the systematic effects mentioned above (uncertainty marked in parenthesis with "sys"), we provide a half-life for the 168 keV isomer of  $T_{1/2}=90(^{+410}_{-50})_{stat}(\pm40)_{tran}(^{+800}_{-70})_{sys}$ ns. The resulting half-life corresponds to 
B(E2)$~=70(^{+80}_{-60})_{stat}(^{+30}_{-40})_{tran}(^{+210}_{-60})_{sys}$ e$^2$fm$^4$.


{\it Discussion.} In the heavy Magnesium isotopes two neutron magic numbers  are
  washed out by the presence of intruder configurations whose energy, fostered by
  the quadrupole correlations and the reductions of the spherical {\it N}=20 and {\it N}=28 gaps, make them dominant in the ground states, giving rise to the {\it N}=20 and
  {\it N}=28 Islands of Inversion. 
  As explained in ref. \cite{Caurier2014} we submit that the effective interaction for this wide region should include a 
  three body monopole correction of the  form:

      \begin{equation}
          \delta V_{pf} = \frac{1}{2}  n_{pf} \displaystyle{(\frac{18}{A})^{1/3}} \;75 \; \rm{keV}
     \end{equation}

    \noindent which restores the {\it N}=20 closure in $^{40}$Mg as it is required by consistency and naturalness of
    the very SM-CI approach ($n_{pf}$ is the number of neutrons in the pf shell in the normal filling approximation).
    Notice however that in ref. \cite{Tsunoda2020}  the {\sl EEdf1} interaction produces a completely different
    picture because  the neutron $0d_{3/2}$ orbit is only half filled already in $^{30}$Mg and remains so even in the $^{40}$Mg ground state.
    With the interaction {\sl {\sl sdpf-u-mix}} the {\it N}=20 neutron gap in $^{40}$Mg is 1 MeV smaller that in $^{32}$Mg, and the
    amount of 2$\hbar \omega$ components in the former is about 25\%. Including the three body correction the effective {\it N}=20 neutron gap
   in $^{40}$Mg increases 2 MeV and the 2$\hbar \omega$ components go down to less than 5\%.

\begin{figure}
\begin{center}
\includegraphics[width=3.4in]{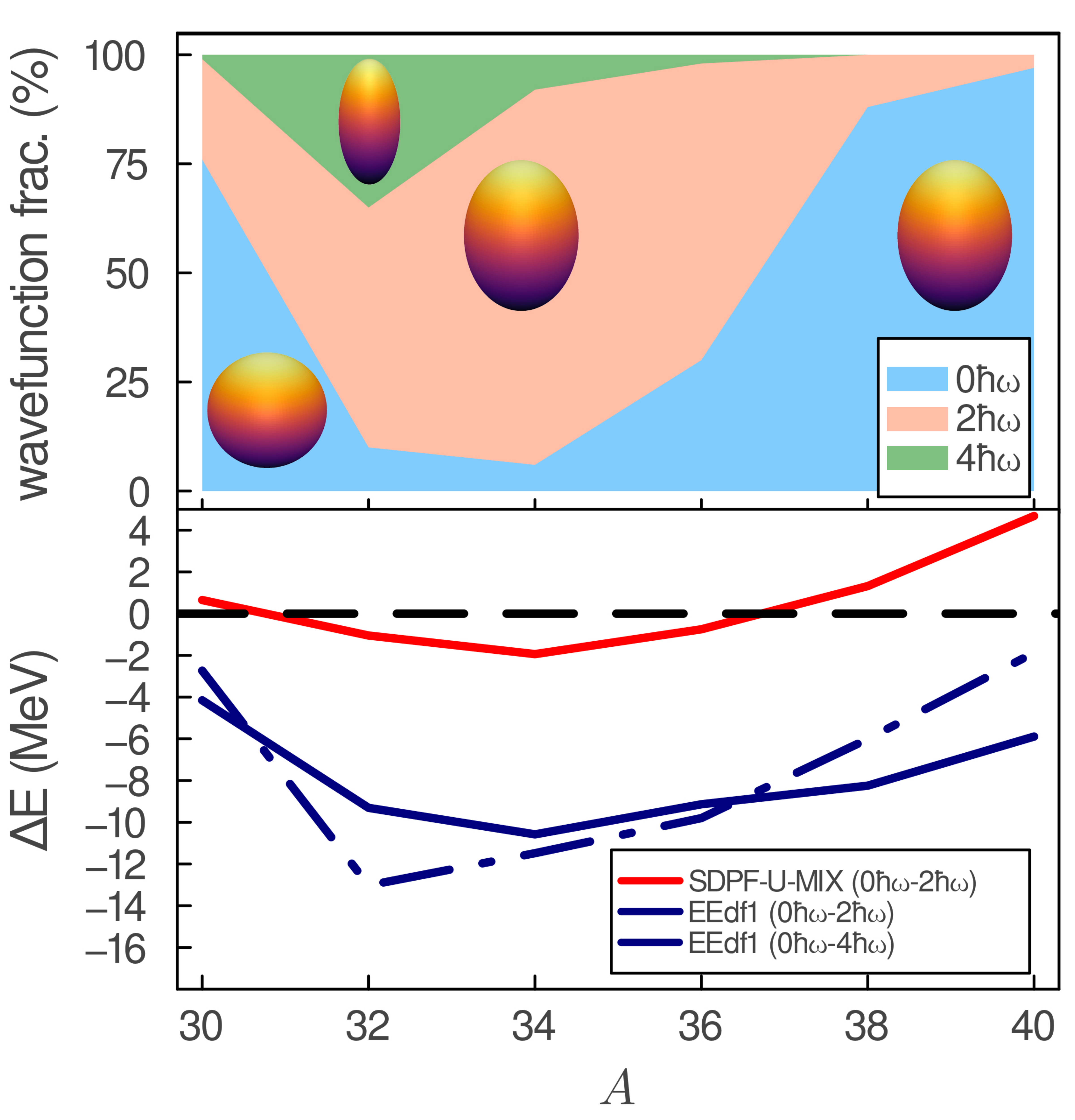}
\caption{(Bottom panel) Energy splitting between the normal filling configuration and the intruder state corresponding to the promotion of two neutrons across
        the {\it N}=20 magic shell closure ($2\hbar\omega$), in the heavy Magnesium isotopes, calculated with {\sl sdpf-u-mix} (red) and {\sl EEdf1} (blue corresponds to $2\hbar\omega$ and dash-dotted blue to 4 neutron excitations $4\hbar\omega$). (Upper panel) Percentage of $x\hbar\omega$ configurations in the theoretical 
        wave functions of the ground states of the Magnesium isotopes, $0\hbar\omega$ (blue), $2\hbar\omega$ (orange), $4\hbar\omega$(green), using the {\sl sdpf-u-mix} interaction. The cartoon nuclei represent the shape of each configuration.  
          }
\label{fig4}
\end{center}
\end{figure}

     We proceed now to explain the results of our SM-CI calculations using the {\sl {\sl sdpf-u-mix}} interaction \cite{Caurier2014}. The calculations
     are performed with the code Antoine \cite{Caurier2005}. It is interesting to follow the location of the 
  intruder configurations before mixing as the number of neutrons increase.
 We denote by $0\hbar\omega$ the normal filling configuration, and by $x\hbar\omega$, with $x=0,2,4$ etc, the 
 {\it x}p-{\it x}h neutron excitations across {\it N}=20. Their relative position gives us a clear
 hint of what will be the structure of the low lying states after full diagonalization,
 although the details of the spectroscopy depend on many other ingredients, in
 particular on the off-diagonal matrix elements between the states calculated
 at fixed values of $x\hbar\omega$ differing in two units.   As a first step, we
     study the relative location of the 0$^+$ band heads of the  $0\hbar\omega$ and $2\hbar\omega$ configurations.  The results are shown in the bottom panel of Fig. \ref{fig4}.
     It is seen that the  intruder states cease to be clearly dominant at {\it A}=36. 
      The increase of the quadrupole moment Q$_{spec}$ of the $2\hbar\omega$ configuration with respect to
      the $0\hbar\omega$ one is small, therefore the gain in quadrupole correlation energy of the latter
      barely compensates its loss of monopole energy. As a consequence, the
      $2\hbar\omega$ configuration in $^{36}$Mg is not as dominant as the one in $^{34}$Mg.  Beyond {\it A}=36 ({\it N}=24), the
      $0\hbar\omega$ configurations are 
      re-established as the main components in the ground states of $^{38}$Mg and $^{40}$Mg. This is in stark contrast with the calculation using the {\sl EEdf1} interaction shown in the bottom panel of Fig. \ref{fig4}. Here, intruder configurations, both $2\hbar\omega$ and $4\hbar\omega$ remain dominant across the entire isotopic chain, precluding mixing with $0\hbar\omega$ normal-filling configurations. However, we must point out, as seen in Fig. \ref{fig4}, the energies of the $2\hbar\omega$ and $4\hbar\omega$ configurations are quasi-degenerate for $^{34}$Mg and $^{36}$Mg. As we will see later, large off-diagonal elements in {\sl EEdf1} result in a strong repulsion when considering two-state mixing, resulting in high energy $0^+_2$ states except for the postulated second 0$^+$ in $^{40}$Mg \cite{Tsunoda2020}. 
      
      

       It is precisely the crossing of the $0\hbar\omega$ and $2\hbar\omega$ configurations in  $^{36}$Mg what might explain the
      very low energy of the proposed 0$^+_2$.   In the full calculation, {\sl {\sl sdpf-u-mix}}  places the excited 0$^+$ in $^{36}$Mg at 1.55 MeV,
       and produces a $^{36}$Mg which is non axially-symmetric (triaxial)  with low energy excitations (gamma band). These low energy states are not compatible with the existence of
       a 0$^+$ isomer as proposed in the present experiment. Given that the amplitude of the $4\hbar\omega$ configurations is negligible,
       the problem can be  translated into a two state model including only the
       $0\hbar\omega$ and the $2\hbar\omega$ states discussed above. According to the calculated Q$_{spec}$ moments within the bands both are prolate deformed.  If the
     energies of the $0\hbar\omega$ and  $2\hbar\omega$ states were degenerate before mixing (and this is nearly the case),  the final splitting 
     of the two 0$^+$ states would be roughly equal   to 2W, with $W = \langle 0\hbar \omega  (0^+) | V  | 2\hbar \omega (0^+) \rangle$.
     In fact,  W=730 keV in $^{36}$Mg, and this sets  a theoretical lower limit to the excitation energy 
      of the isomer 0$^+$, within the two state model. The value  in the complete diagonalization 
      is very much in line with this estimate. Thus, the only way to get  the splitting right
      is via a reduction of the value of W, which is dominated by the off-diagonal pairing interaction between the
      sd and the pf-shell neutron orbits. Hence we are led to make an “ad hoc”  10\% reduction of the off diagonal pairing matrix elements for $n_{pf} >$ 0,
      bringing W down to about 500 keV.
     With this choice, the resulting composition of the ground states of the Magnesium isotopes is as depicted in the upper panel of Figure \ref{fig4}. We see that the intruder ($2\hbar \omega,4\hbar\omega$) configurations are dominant in $^{32,34,36}$Mg, while the normal-filling ($0\hbar\omega$) states take the majority of the wavefunction in $^{30,38,40}$Mg. This trend is consistent with the restoration of the {\it N}=20 shell closure as we approach the {\it N}=28 subshell, disfavoring particle-hole excitations across the {\it N}=20 shell gap.

    The spectroscopic results for $^{36}$Mg are gathered in Table \ref{tab:mg36}.
\begin{table}[h]
\caption{\label{tab:mg36} Theoretical excitation energies (in MeV), Q$_{spec}$ in $efm^2$ and B(E2)’s (in $e^2fm^4$), for $^{36}$Mg, using the {\sl sdpf-u-mix} interaction.}
\begin{tabular*}{\linewidth}{@{\extracolsep{\fill}}|ccccc|}
\hline  
J$^{\pi}$ &  E(th))&  Q$_{spec}$ &  J$^{\pi}$(f) & B(E2) \\ 
\hline
$0^+_1$  & 0.0 &          &      &        \\ 
$2^+_1$  & 0.58 &   -23      &  $0^+_1$  & 130 \\ 
$0^+_2$  & 1.02 &        &   $2^+_1$  & 5  \\ 
$2^+_2$  & 1.43 &  -15      &   $0^+_1$   &  2 \\ 
$2^+_2$  &         &        &   $2^+_1$   &  1 \\ 
$2^+_2$  &         &         &  $0^+_2$   & 120 \\ 
$4^+_1$  & 1.73 &   -23     &   $2^+_1$   & 183 \\ 
$4^+_1$  &         &        &   $2^+_2$   & 1 \\ 
\hline    
\end{tabular*}
\end{table}   
 The excitation energies are in good agreement  with the present experimental result for the proposed  0$^+$ isomer and
        with the results of reference \cite{Doornenbal2013} for the yrast 2$^+$ and  4$^+$. Using the {\sl EEdf1} interaction in Monte Carlo Shell Model, we obtained a 0$^+_2$ energy of 2.32 MeV with a B(E2) of 0.4 e$^2$fm$^4$, using effective charges of 1.25 e and 0.25 e for protons and neutrons respectively, not compatible with our observation. 
               We must stress that the possible presence of a $0^+$ isomer in $^{36}$Mg is important beyond the value of its excitation energy.
        As mentioned above, our calculations with the {\sl sdpf-u-mix} interaction show different  
        nuclear structure depending on the energy of the $0^+_2$ state, from a triaxial solution if it were not isomeric,  to a case of
        two coexisting prolate bands. Figure \ref{fig4} and the E2 properties listed in table \ref{tab:mg36}  show the lowest band is dominated by $2\hbar\omega$ configurations and the excited one by $0\hbar\omega$ configurations. A very prominent feature of them is that the configuration mixing between the two
        bands is almost absent. In particular, the B(E2) from the isomer to the yrast  2$^+$     is small,  compatible with the
        experimental value extracted from its lifetime.
        Let’s mention finally that the crossover from the dominance of the $0f_{7/2}$ orbit to a massive occupation of
        the $1p_{3/2}$  orbit, takes place at {\it N}=24 as well, paving the way to the {\it N}=28 Island of Inversion.

{\it Conclusions.} We observed a new 168 keV isomeric transition in $^{36}$Mg, with a half-life of $90(^{+410}_{-50})_{stat}(\pm40)_{tran}(^{+800}_{-70})_{sys}$ ns with $3\sigma$ statistical and systematic uncertainties. From the observation of a 665 keV gamma line in the prompt gamma spectrum, we propose that it corresponds to a new $0^+$ state at  833 keV de-exciting to the known 665 keV $2^+$ state \cite{Doornenbal2016}. To elucidate the microscopic origin of this isomer we performed shell model calculations using the {\sl sdpf-u-mix} interaction. We propose the observed low excitation energy of the state arises from two coexisting prolate deformed configurations consisting of the normal-filling and intruder two-neutron excitations respectively. We predict that, for N$>$20 Mg isotopes, as the neutron $0f_{7/2}$ orbital is gradually filled, the {\it N}=20 shell closure is restored while the {\it N}=28 subshell closure is quenched. Therefore, the quasi-degeneracy between normal and intruder configurations occurs only for $^{36}$Mg. In contrast, other effective interactions used so far in the region predict substantial quenching of the {\it N}=20 shell closure even past $^{38}$Mg. We postulate the discrepancy arises from the inclusion, or lack of thereof, of three-body corrections into the monopole part of the effective interaction. The isomer presented in this work supports that $^{36}$Mg is the bridge between the {\it N}=20 island of inversion centered around $^{32}$Mg and the {\it N}=28 island of inversion centered in $^{40}$Mg. As a direct consequence we anticipate no isomers will be present in $^{34,38}$Mg. Thanks to the large yields of magnesium isotopes afforded at the recently commissioned  FRIB facility in MSU (or RIKEN, Japan), this hypothesis may be tested in the near future. 
We thank the NSCL operations staff for providing the excellent 
quality radioactive ion beams necessary for this work.
\begin{acknowledgments}
{\it Acknowledgments.} We thank Augusto Macchiavelli for the fruitful discussions during the preparation of this manuscript. The isotope(s) used in this research was supplied by the Isotope Program within the Office of Nuclear Physics in the Department of Energy’s Office of Science. The Lanczos shell-model calculation using the {\sl EEdf1} interaction was performed with the code KSHELL \cite{shimizu2019}. This research was sponsored in part by the National Nuclear Security Administration under the Stewardship Science Academic Alliances program through DOE Cooperative Agreements No. DE-NA0003899 and  DE-NA0004068.
This research was also sponsored by the Office of Nuclear Physics, U. S. Department 
of Energy under contract DE-FG02-96ER40983 (UT) and DE-SC0020451 (MSU). This research was sponsored in part by  National Science Foundation under the contract NSF-MRI-1919735. This work was also supported by the National Nuclear Security Administration through the Nuclear Science and Security Consortium under Award No. DE-NA0003180 and the Stewardship Science Academic Alliances program through DOE Award No. DOE-DE-NA0003906. N.S. and Y.U. acknowledge computer resources provided by U. Tsukuba MCRP program (woi22i022) and "Program for Promoting Researches on the Supercomputer Fugaku"
(JPMXP1020200105).
A.P. is supported by
grants CEX2020-001007-S  funded by MCIN/AEI/10.13039/501100011033 and PID2021-127890NB-I00.
\end{acknowledgments}


%

\end{document}